\documentclass[preprint,prb,amsmath]{revtex4}
\usepackage{bm}
\usepackage{graphicx}

\input{tcilatex}
\begin{document}

\author{W. J. Mullin$^{a}$, R. Krotkov$^{a}$, and F. Lalo\"{e}$^{b}$}
\title{The origin of phase in the interference of Bose-Einstein condensates}
\affiliation{$^{a}$Department of Physics, University of Massachusetts,
Amherst, MA 01003 USA\\
$^{b}$LBK, Dept.\ de Physique de l'ENS, 24 rue Lhomond, 75005, Paris, France}
\date{\today}

\begin{abstract}
We consider the interference of two overlapping ideal Bose-Einstein
condensates. The usual description of this phenomenon involves the
introduction of a so-called condensate wave functions having a definite
phase. We investigate the origin of this phase and the theoretical basis of
treating interference. It is possible to construct a phase state, for which
the particle number is uncertain, but phase is known. However, how one would
prepare such a state before an experiment is not obvious. We show that a
phase can also arise from experiments using condensates in Fock states, that
is, having known particle numbers. Analysis of measurements in such states
also gives us a prescription for preparing phase states. The connection of
this procedure to questions of ``spontaneously broken gauge symmetry'' and to
``hidden variables'' is mentioned.
\end{abstract}

\maketitle


\section{INTRODUCTION}

\label{sec:Intro} One of the most impressive experiments using trapped Bose
gases is the interference experiment of Ketterle and co-workers.\cite
{Ketterle} Two condensates are are separately prepared and allowed to
overlap. An interference pattern then arises showing the remarkable quantum
coherence of the condensates. There have been other interesting condensate
interference experiments as well.\cite{Int2}$^{-}$\cite{Hall} If one assumes
that the separated clouds initially have a definite phase relation, then the
experiments are well-described by straight-forward theory.\cite{Nara}
However, questions immediately arise. Do separately prepared condensates
have a phase relation?\cite{A-3} The preparation of the sample certainly did
not involve the establishment of a state with known phase. More likely the
particle number in each cloud would have, or could have, been initially
known ahead of time. Nevertheless an interference pattern emerges with some
well-established phase. So how is this possible? The question was answered
in several theoretical papers showing how a phase appears even when the two
clouds are prepared in Fock states, that is, states with sharply known
particle numbers.\cite{JH}$^{-}$\cite{PS} In this paper we revisit the
question and present a derivation of this result. This result is quite
satisfying, since it in a sense justifies the usual simple assumption of the
interferring coherent systems having a well-defined, but completely unknown,
initial phase relationship.

In order to discuss the properties of condensates and superfluids one
usually introduces theoretically a so-called ``order parameter'' or
``condensate wave function'' defined as $\left\langle \hat{\psi}(\mathbf{r}%
)\right\rangle $ where $\hat{\psi}(\mathbf{r})$ is an operator destroying a
particle at position $\mathbf{r.}$ The resulting quantity has a magnitude
equal to the square root of the condensate density and also a phase.
Obviously the state in which $\left\langle \hat{\psi}(\mathbf{r}%
)\right\rangle $ is non-zero cannot have a fixed particle number. With one
of these wave functions for each condensate, it is straightforward to
discuss interference of the two, since each one has its own phase, and an
interference pattern arises with a relative phase equal to the difference
between the individual phases. In essence one has described each condensate
by a single-particle wave function so that interference is no more than the
overlap and interference of two classical waves. However, one can ask how
this single-particle wave function might have arisen. Indeed its existence
involves the question of ``spontaneously broken gauge symmetry''\cite{A-2}$%
^{-}$\cite{Leggett}, the necessity of which has been brought into question
in recent years.\cite{LS}$^{-}$\cite{Johnston}

Suppose we consider a condensate described by a wave function $e^{i(\mathbf{k%
\cdot r}+\phi )}.$ One might describe the direction specified by the angle $%
\phi $ by a ``spin'' in a two-dimensional plane. How do we prepare such a
state? What is it that picks out the direction of this pseudo-spin from the
all the degenerate possible directions? There is an analogy with
ferromagnetism, where there is symmetry in the possible degenerate
three-dimensional directions of the magnetization. Somehow, say, because of
a small external field, a particuliar direction is space is picked out for
the magnetization. In a similar way the phase angle is picked out. One can
suppose in the ferromagnetism case that, in actual practice, there is always
some small field to pick out a preferred direction, so that the symmetry is
broken. However, the local field that is used theoretically to pick a phase
direction does not exist in nature. Indeed one would have to prepare a state
that does not have a fixed particle number. The treatment of phase (actually
relative phase) in Sec. IV, and the spontaneous appearance of a relative
phase under the effect of measurement of particle position in Fock states,
avoids violating particle conservation and does not require use of any
symmetry-breaking field and so helps in this regard.

A closely related feature of the discussion is the idea that the phase
emerging from successive measurements of particle position starting with a
Fock state is somewhat like the emergence of a ``hidden'' or ``additional''
variable in quantum mechanics.\cite{FLHidden}$^{,}$\cite{Bell-0}$^{,}$\cite
{FL-1} Was the phase there before the experiment started, or did the
experiment itself cause it to take on its final value? Hidden variables can
be invoked to specify non-commuting variables. In standard quantum
mechanics, particle number and relative phase can be considered conjugate
variables; the knowledge of one excludes that of the other. As we measure
particle position we will see that the knowledge of the particle number
becomes less certain while the uncertainty of relative phase decreases.

In the next sections we first discuss the kind of state that has a known
phase. Obviously with this state the interference pattern emerges with just
the prepared phase. Among these states are are the coherent states of
Glauber.\cite{G} These can either have particle number completely unknown or
have the total number of particles in two clouds known (in which case they
are called ``phase states''), although the number in each cloud is still
unknown. We then derive the interference pattern starting with Fock states
and see the emergence of a relative phase even though no phase was present
at the beginning of the experiment (or was at least ``hidden''). We even
find a way to prepare a state that has a known relative phase. The
controversal theoretical constructs are seen to be unnecessary.

\section{SIMPLE VIEW OF AN INTERFERENCE PATTERN}

\label{sec:repulsion}A gas (or liquid) undergoing Bose-Einstein condensation
(BEC) is often described by a classical field known as an ``order
parameter'' or ``condensate wave function.'' Such a quantity can arise in
several ways. Suppose that $\hat{\psi}(\mathbf{r})$ represents a
second-quantized operator destroying a boson at position $\mathbf{r}$. Then
the one-particle density matrix is defined as $\rho (\mathbf{r},\mathbf{r}%
^{\prime })=\left\langle \hat{\psi}^{\dagger }(\mathbf{r})\hat{\psi}(\mathbf{%
r})\right\rangle .$ Penrose and Onsager\cite{PO} showed that a criterion for
a Bose condensate or ``off-diagonal long-range order'' is that the density
matrix have the form 
\begin{equation}
\rho (\mathbf{r},\mathbf{r}^{\prime })=\psi ^{*}(\mathbf{r})\psi (\mathbf{r}%
^{\prime })+f(\mathbf{r},\mathbf{r}^{\prime })
\end{equation}
where $f(\mathbf{r},\mathbf{r}^{\prime })$ vanishes when $\left| \mathbf{r}-%
\mathbf{r}^{\prime }\right| \rightarrow \infty $ . The function $\psi (%
\mathbf{r})$ is the condensate wave function. One often assumes the system
in a state such that the destruction operator itself has a non-zero average: 
\begin{equation}
\left\langle \hat{\psi}(\mathbf{r})\right\rangle =\psi (\mathbf{r})=\sqrt{%
n_{0}(\mathbf{r})}e^{i\phi (\mathbf{r})}  \label{orderpar}
\end{equation}
where $n_{0}$ is the condensate density and $\phi $ its phase. Such a state
is said to have ``spontaneously broken gauge symmetry'' because a particular
phase (out of many possible degenerate phase states) has been chosen.\cite
{A-2}$^{-}$ \cite{Johnston} The Gross-Pitaevskii equation is a non-linear
Schrodinger equation for $\psi (\mathbf{r}),$ which has been used
extensively, with remarkable success, to describe interacting trapped BEC
gases within mean-field approximation.\cite{Stringari}

In describing the interference pattern of the experiment of Ref.\ %
\onlinecite {Ketterle} one has to consider the overlap of Bose clouds
released from harmonic oscillator traps.\cite{Nara} This leads to some
interesting features such as fringes whose separation changes with time. In
our analysis here we will consider only plane waves and ignore any time
evolution. Thus suppose we have an order parameter that involves two
condensate clouds, having condensate densities $n_{a}$ and $n_{b}$ in
momentum states $\mathbf{k}_{a}$ and $\mathbf{k}_{b}.$ This dual order
parameter has the form 
\begin{equation}
\psi (\mathbf{r})=\sqrt{n_{a}}e^{i\mathbf{k}_{a}\mathbf{r}}e^{i\phi _{a}}+%
\sqrt{n_{b}}e^{i\mathbf{k}_{b}\mathbf{r}}e^{i\phi _{b}}.  \label{simplecase}
\end{equation}
The density of the combined system is then 
\begin{equation}
n(\mathbf{r})=\left| \sqrt{n_{a}}e^{i\mathbf{k}_{a}\mathbf{r}}e^{i\phi _{a}}+%
\sqrt{n_{b}}e^{i\mathbf{k}_{b}\mathbf{r}}e^{i\phi _{b}}\right| ^{2}=n\left(
1+x\cos (\mathbf{k\cdot r}+\phi )\right)   \label{SimpInt}
\end{equation}
where $n=n_{a}+n_{b},$ $\mathbf{k}=\mathbf{k}_{a}-\mathbf{k}_{b}$, $\phi
=\phi _{b}-\phi _{a},$ and $x=2\sqrt{n_{a}n_{b}}/n.$ We have an interference
pattern with \emph{relative} phase $\mathbf{k\cdot r}+\phi .$ The phase
shift $\phi $ is measureable although the individual phases $\phi _{a}$ and $%
\phi _{b}$ are not.

This analysis is simple, but it requires the preparation of the system in a
state with known individual phases. How do we do that? What is the nature of
such a state? Clearly the expectation value of Eq.\ (\ref{orderpar}) cannot
be in a state of definite particle number or the expectation value would
vanish. Next we investigate this question more deeply.

\section{PHASE STATES}

As noted by Johnston\cite{Johnston} in this journal, the coherent states
introduced by Glauber\cite{G} for photons are appropriate for superfluids.
These well-know states have been reviewed in this journal on occasion,\cite
{coherent} and appear in texts\cite{CTDL} as well. They are also called
``classical states'' and are the minimum uncertainty states of the harmonic
oscillator.\cite{CTDL} Here we do not use them in full generality, but
rather use a subset of them known as ``phase states.'' (Those wishing to see
the full treatment of coherent states in a treatment of a condensate wave
function are referred to Appendix A.) Phase states describe two condensates
(in states $\mathbf{k}_{a}$ and $\mathbf{k}_{b}$) with variable particle
numbers, $N_{a},$ $N_{b}$ (both macroscopic), but fixed total number $%
N=N_{a}+N_{b}.$ No other momentum states are occupied. If particle creation
operators $a^{\dagger }$ and $b^{\dagger }$ (obeying Bose commutation
relations) for the two states act on the vacuum to put particles into these
two states, then we define the (properly normalized) state as 
\begin{eqnarray}
\left| \alpha _{a}\alpha _{b};N\right\rangle  &=&\frac{1}{\sqrt{g^{N}}}%
\sum_{N_{a}=0}^{N}\sqrt{\frac{N!}{N_{a}!\left( N-N_{a}\right) !}}\alpha
_{a}^{N_{a}}\alpha _{b}^{N-N_{a}}\left| N_{a},N-N_{a}\right\rangle  
\nonumber \\
&=&\frac{1}{\sqrt{g^{N}}}\sum_{N_{a}}\frac{\sqrt{N!}}{N_{a}!\left(
N-N_{a}\right) !}\left( \alpha _{a}a^{\dagger }\right) ^{N_{a}}\left( \alpha
_{b}b^{\dagger }\right) ^{N-N_{a}}\left| 0\right\rangle   \nonumber \\
&=&\sqrt{\frac{1}{g^{N}N!}}\left( \alpha _{a}a^{\dagger }+\alpha
_{b}b^{\dagger }\right) ^{N}\left| 0\right\rangle   \label{Expform}
\end{eqnarray}
where we define the quantities $\alpha _{i}$ as complex and separate them
into magnitudes $\gamma _{i}$ and phases $\phi _{i}$ according to the
notation 
\begin{equation}
\alpha _{i}=\gamma _{i}e^{i\phi _{i}}.
\end{equation}
Also $g=\sqrt{\gamma _{a}^{2}+\gamma _{b}^{2}}.$ We can easily compute the
average number of particles $\bar{N}_{a}$ in this state$.$ Use the first
form of Eq. (\ref{Expform}) to give 
\begin{eqnarray}
a\left| \alpha _{a}\alpha _{b};N\right\rangle  &=&\frac{1}{\sqrt{g^{N}}}%
\sum_{N_{a}}\sqrt{\frac{N!}{N_{a}!\left( N-N_{a}\right) !}}\alpha
_{a}^{N_{a}}\alpha _{b}^{N-N_{a}}\sqrt{N_{a}}\left|
N_{a}-1,N-N_{a}\right\rangle   \nonumber \\
&=&\alpha _{a}\frac{\sqrt{N}}{\sqrt{g^{N}}}\sum_{N_{a}^{\prime }}\sqrt{\frac{%
(N-1)!}{N_{a}^{\prime }!\left( N-1-N_{a}^{\prime }\right) !}}\alpha
_{a}^{N_{a}^{\prime }}\alpha _{b}^{N-1-N_{a}^{\prime }}\left| N_{a}^{\prime
},N-N_{a}\right\rangle   \nonumber \\
&=&\alpha _{a}\sqrt{\frac{N}{g}}\left| \alpha _{a}\alpha
_{b};N-1\right\rangle 
\end{eqnarray}
where we have taken $N_{a}^{\prime }=N_{a}-1.$ Thus 
\begin{equation}
\bar{N}_{a}=\left\langle \alpha _{a}\alpha _{b};N\right| a^{\dagger }a\left|
\alpha _{a}\alpha _{b};N\right\rangle =\gamma _{a}^{2}\frac{N}{\gamma
_{a}^{2}+\gamma _{b}^{2}}
\end{equation}
Similarly we get $\bar{N}_{b}=\gamma _{b}^{2}N/(\gamma _{a}^{2}+\gamma
_{b}^{2})$ so that $\gamma _{i}=\left| \alpha _{i}\right| =\sqrt{\bar{N}_{i}}
$ and $g=N_{a}+N_{b}=N.$

The fact that $N$ is known in our phase state does not affect the results
for interference patterns, which depend just on relative phase. Indeed such
states have been used on many occasions to discuss the interference of two
condensates.\cite{CD}$^{,}$\cite{HB}$^{,}$\cite{PS}$^{,}$\cite{KS} Our state
can be used to discuss the how relative phase can be conjugate to particle
number. Write it in a form that makes the phases explicit: 
\begin{equation}
\left| \alpha _{a}\alpha _{b};N\right\rangle =\sqrt{\frac{N!}{g^{N}}}%
\sum_{\left( N_{a}+N_{b}=N\right) }\frac{\gamma _{a}^{N_{a}}\gamma
_{b}^{N_{b}}e^{iN_{a}\phi _{a}}e^{iN_{b}\phi _{b}}}{\sqrt{N_{a}!N_{b}!}}%
\left| N_{a},N_{b}\right\rangle
\end{equation}
Now change the phases to relative phase $\phi =\phi _{b}-\phi _{a}$ and
total phase $\Phi =(1/2)(\phi _{b}+\phi _{a})$ and then take the derivative
with respect to $\phi :$%
\begin{eqnarray}
-2i\frac{\partial }{\partial \phi }\left| \alpha _{a}\alpha
_{b};N\right\rangle &=&\sqrt{\frac{N!}{g^{N}}}\sum_{\left(
N_{a}+N_{b}=N\right) }(N_{a}-N_{b})\frac{\gamma _{a}^{N_{a}}\gamma
_{b}^{N_{b}}e^{iN_{a}(\Phi +\phi /2)}e^{iN_{b}(\Phi -\phi /2)}}{\sqrt{%
N_{a}!N_{b}!}}\left| N_{a},N_{b}\right\rangle  \nonumber \\
&=&\left( a^{\dagger }a-b^{\dagger }b\right) \left| \alpha _{a}\alpha
_{b};N\right\rangle
\end{eqnarray}
The phase derivative operator gives the same result as the number difference
operator so that $\phi $ and $(N_{a}-N_{b})$ are conjugate variables.\cite
{Leggett}$^{,}$\cite{KS} Note that the total phase appears only as an
external factor $\exp \left[ iN\Phi \right] $ and so has no physical
significance. Thus the individual phases have no physical significance; only
the relative phase $\phi $ is a meaningful quantity.

In order to emphasize this last point and put the phase state in a more
compact form to treat interference we rename it and rewrite it as 
\begin{equation}
\left| \phi ,N\right\rangle =\frac{1}{\sqrt{g^{N}N!}}(a^{\dagger }+\gamma
e^{i\phi }b^{\dagger })^{N}\left| 0\right\rangle =\frac{1}{\sqrt{g^{N}}}%
\sum_{n=1}^{N}\sqrt{\frac{N!}{n!(N-n)!}}(\gamma e^{i\phi })^{N-n}\left|
n,N-n\right\rangle   \label{phasestate}
\end{equation}
where now $\gamma =\sqrt{\bar{N}_{a}/\bar{N}_{b}}$ and $\phi $ is the
relative phase. Also now $g=(1+\gamma ^{2}).$ We have simply dropped a
meaningless factor of unit magnitude.

Since we have just two occupied states, the terms in Eq.\ (\ref{field exp})
not referring to states $k_{a}$ and $k_{b}$ never contribute and we can more
simply write 
\begin{equation}
\hat{\psi}(\mathbf{r})\rightarrow c_{\mathbf{r}}\equiv \sqrt{\frac{1}{V}}%
\left( e^{i\mathbf{k}_{a}\cdot \mathbf{r}}a+e^{i\mathbf{k}_{b}\cdot \mathbf{r%
}}b\right) \mathbf{.}
\end{equation}
We can also make this more compact by writing 
\begin{equation}
c_{\mathbf{r}}=\sqrt{\frac{1}{V}}\left( a+e^{i\mathbf{k}\cdot \mathbf{r}%
}b\right) 
\end{equation}
with $\mathbf{k}=\mathbf{k}_{b}-\mathbf{k}_{a}.$ The quantity extracted out $%
e^{i\mathbf{k}_{a}\cdot \mathbf{r}}$ can again be dropped as a meaningless
unit-magnitude factor. We want to consider how $c_{\mathbf{r}}$ acts on the
phase state. We have 
\begin{eqnarray}
c_{\mathbf{r}}\left| \phi ,N\right\rangle  &=&\frac{1}{\sqrt{g^{N}V}}%
\sum_{n=1}^{N}\sqrt{\frac{N!}{n!(N-n)!}}(\gamma e^{i\phi })^{N-n} \\
&&\times \left( \sqrt{n}\left| n-1,N-n\right\rangle +e^{i\mathbf{k}\cdot 
\mathbf{r}}\sqrt{N-n}\left| n,N-n-1\right\rangle \right) 
\end{eqnarray}
By changing variables in the first state to $n^{\prime }=n-1$ we can put
both terms in the same form so 
\begin{equation}
c_{\mathbf{r}}\left| \phi ,N\right\rangle =A(\mathbf{r,}\phi )\left| \phi
,N-1\right\rangle   \label{eigen}
\end{equation}
where 
\begin{equation}
A(\mathbf{r,}\phi )=\sqrt{\frac{N}{gV}}\left( 1+\gamma e^{i\phi }e^{i\mathbf{%
k}\cdot \mathbf{r}}\right) 
\end{equation}
We can now easily compute the condensate wave function as 
\begin{equation}
\left\langle \phi ,N\right| \hat{\psi}(\mathbf{r})\left| \phi
,N\right\rangle =\left\langle \phi ,N\right| c_{\mathbf{r}}\left| \phi
,N\right\rangle =A(\mathbf{r,}\phi )
\end{equation}
If for the moment we put back the previously neglected leading phase factors
we have 
\begin{equation}
\left\langle \phi ,N\right| \hat{\psi}(\mathbf{r})\left| \phi
,N\right\rangle =\sqrt{\bar{n}_{a}}e^{i\mathbf{k}_{a}\cdot \mathbf{r}%
}e^{i\phi _{a}}+\sqrt{\bar{n}_{b}}e^{i\mathbf{k}_{b}\cdot \mathbf{r}%
}e^{i\phi _{b}}
\end{equation}
which has just the same form as Eq.\ (\ref{simplecase}), with average
densities in place of precise densities and well-defined individual phases.
However the individual phases are not measurable---only the relative phase
is. Indeed the average density follows immediately as 
\begin{equation}
\left\langle \phi ,N\right| \hat{\psi}^{\dagger }(\mathbf{r})\hat{\psi}(%
\mathbf{r})\left| \phi ,N\right\rangle =\left\langle \phi ,N\right| c_{%
\mathbf{r}}^{\dagger }c_{\mathbf{r}}\left| \phi ,N\right\rangle =\left| A(%
\mathbf{r},\phi )\right| ^{2}=\bar{n}\left( 1+\bar{x}\cos (\mathbf{k\cdot r}%
+\phi )\right) 
\end{equation}
just like Eq.\ (\ref{SimpInt}) where now $\bar{n}$ and $\bar{x}$ have
obvious definitions in terms of averages. Thus a phase state provides a
rigorous background for discussion of condensate wave functions and for the
simplified form of treating interference between the two condensates. How we
might actually prepare one \emph{before} the experiment is a separate
difficult question, which we treat below.

We will find it useful and necessary in the next section to consider more
general cases in which we make measurements of many particle positions
essentially simultaneously. This allows interference fringes to emerge where
they would otherwise not occur. For our phase state such calculations are
straightforward and add no additional information since the the
multiparticle densities all factor in the phase states. For example,
consider the expectation value of 
\begin{equation}
\hat{\psi}^{\dagger }(\mathbf{r}_{2})\hat{\psi}(\mathbf{r}_{2})\hat{\psi}%
^{\dagger }(\mathbf{r}_{1})\hat{\psi}(\mathbf{r}_{1})=c_{\mathbf{r}%
_{2}}^{\dagger }c_{\mathbf{r}_{2}}c_{\mathbf{r}_{1}}^{\dagger }c_{\mathbf{r}%
_{1}}\approx c_{\mathbf{r}_{2}}^{\dagger }c_{\mathbf{r}_{1}}^{\dagger }c_{%
\mathbf{r}_{2}}c_{\mathbf{r}_{1}}.
\end{equation}
Of course, $c_{\mathbf{r}_{2}}$ and $c_{\mathbf{r}_{1}}^{\dagger }$ don't
commute, but in the approximation shown we are dropping a term of order $N$
compared to one of order $N^{2}.$ The last form is more convenient to use.
By the eigenfunction behavior of the phase state we easily get 
\begin{eqnarray}
\left\langle \phi ,N\right| \hat{\psi}^{\dagger }(\mathbf{r}_{2})\hat{\psi}(%
\mathbf{r}_{2})\hat{\psi}^{\dagger }(\mathbf{r}_{1})\hat{\psi}(\mathbf{r}%
_{1})\left| \phi ,N\right\rangle  &\approx &\left\langle \phi ,N\right| c_{%
\mathbf{r}_{2}}^{\dagger }c_{\mathbf{r}_{1}}^{\dagger }c_{\mathbf{r}_{2}}c_{%
\mathbf{r}_{1}}\left| \phi ,N\right\rangle =\left| A(\mathbf{r}_{2})\right|
^{2}\left| A(\mathbf{r}_{1})\right| ^{2}  \nonumber \\
&=&\prod_{i=1}^{2}\bar{n}\left( 1+\bar{x}\cos (\mathbf{k\cdot r}_{i}+\phi
)\right)   \label{2bodyphase}
\end{eqnarray}
This result generalizes to 
\begin{eqnarray}
\left\langle \phi ,N\right| \hat{\psi}^{\dagger }(\mathbf{r}_{m})\hat{\psi}(%
\mathbf{r}_{m})\cdots \hat{\psi}^{\dagger }(\mathbf{r}_{1})\hat{\psi}(%
\mathbf{r}_{1})\left| \phi ,N\right\rangle  &\approx &\left\langle \phi
,N\right| c_{\mathbf{r}_{m}}^{\dagger }\cdots c_{\mathbf{r}_{1}}^{\dagger
}c_{\mathbf{r}_{m}}\cdots c_{\mathbf{r}_{1}}\left| \phi ,N\right\rangle  
\nonumber \\
&=&\prod_{i=1}^{m}\bar{n}\left( 1+\bar{x}\cos (\mathbf{k\cdot r}_{i}+\phi
)\right)   \label{mPhase}
\end{eqnarray}

Considering a state in the form $c_{\mathbf{r}_{m}}\cdots c_{\mathbf{r}%
_{1}}\left| \Psi \right\rangle $ is useful for interpreting an experiment.
Here we can consider our experiment as detecting particle 1 and then 2
shortly thereafter, and so on. After $m$ detections the wave function
evolves to a state missing several particles. What is the nature of the
state to which it has evolved? In the case of a phase state it is $c_{%
\mathbf{r}_{m}}\cdots c_{\mathbf{r}_{1}}\left| \phi ,N\right\rangle \sim
\left| \phi ,N-m\right\rangle .$ However, it is more interesting to consider
the case of $\left| \Psi \right\rangle $ a Fock state as we do in the next
section.

\section{INTERFERENCE IN FOCK STATES}

\label{sec:BADIDEA}It is not evident that experimentalists can prepare a
phase state as described above. It would seem more likely that they are
working with Fock states, that is, states in which the particles numbers, $%
N_{a}$ and $N_{b},$ in the two condensates are known rather well. It seems
likely, in any case, that one has more chance of initially preparing such a
state. However, as several workers\cite{JH}$^{-}$\cite{PS} have realized in
recent years, and as we will show, that an interference pattern with some
phase can still arise in a Fock state. Denoting the state sharp in particle
number as $\left| N_{a},N_{b}\right\rangle $ and using the Bose annihilation
relation 
\begin{equation}
c_{\mathbf{r}}\left| N_{a},N_{b}\right\rangle =\sqrt{\frac{1}{V}}\left( 
\sqrt{N_{a}}\left| N_{a}-1,N_{b}\right\rangle +e^{i\mathbf{k}\cdot \mathbf{r}%
}\sqrt{N_{b}}\left| N_{a},N_{b}-1\right\rangle \right) 
\end{equation}
the one-body density in a Fock state is 
\begin{equation}
D_{1}=\left\langle N_{a},N_{b}\right| c_{\mathbf{r}}^{\dagger }c_{\mathbf{r}%
}\left| N_{a},N_{b}\right\rangle =n
\end{equation}
and of course there is no interference. Phase and particle number are
conjugates, and the particle number is initially known.

However, if we consider measuring the position of two particles
simultaneously, some correlation should arise. We have 
\begin{eqnarray}
c_{\mathbf{r}_{2}}c_{\mathbf{r}_{1}}\left| N_{a},N_{b}\right\rangle  &=&%
\frac{1}{V}\left[ \sqrt{N_{a}(N_{a}-1)}\left| N_{a}-2,N_{b}\right\rangle
+e^{i\mathbf{k}\cdot \left( \mathbf{r}_{1}+\mathbf{r}_{2}\right) }\sqrt{%
N_{b}(N_{b}-1)}\left| N_{a},N_{b}-2\right\rangle \right.   \nonumber \\
&&+\left. \sqrt{N_{a}N_{b}}\left( e^{i\left( \mathbf{k}\cdot \mathbf{r}%
_{1}\right) }+e^{i\left( \mathbf{k}\cdot \mathbf{r}_{2}\right) }\right)
\left| (N_{a}-1)(N_{b}-1)\right\rangle \right]   \label{Nlesscert}
\end{eqnarray}
so that the particle number is now slightly less certain.  The
two-body Fock correlation function is
\begin{eqnarray}
D_{2} &=&\left\langle N_{a},N_{b}\right| c_{\mathbf{r}_{1}}^{\dagger }c_{%
\mathbf{r}_{2}}^{\dagger }c_{\mathbf{r}_{2}}c_{\mathbf{r}_{1}}\left|
N_{a},N_{b}\right\rangle   \nonumber \\
&=&\frac{1}{V^{2}}\left[ N_{a}(N_{a}-1)+N_{b}(N_{b}-1)+N_{a}N_{b}\left|
e^{i\left( \mathbf{k}\cdot \mathbf{r}_{1}\right) }+e^{i\left( \mathbf{k}%
\cdot \mathbf{r}_{2}\right) }\right| ^{2}\right]   \nonumber \\
&=&n^{2}\left[ 1+x\cos \mathbf{k\cdot }(\mathbf{r}_{1}-\mathbf{r}%
_{2})\right]   \label{twopart}
\end{eqnarray}
For two particles there is indeed a position correlation.  As we make
more and more measurements the state gets more and more mixed among
states with various numbers of particles.

We can rewrite Eq.\ (\ref{twopart}) in a somewhat different and useful way.
A simple integration shows that 
\begin{equation}
D_{2}=n^{2}\int_{0}^{2\pi }\frac{d\phi }{2\pi }\prod_{i=1}^{2}\left[ 1+x\cos
(\mathbf{k\cdot r}_{i}+\phi )\right] .
\end{equation}
Compare this with Eq.\ (\ref{2bodyphase}). We have a similar looking result
except that we now integrate over all relative phases. Remarkably this
result can be extended to higher order correlation functions. Indeed Ref.\ %
\onlinecite{FLHidden} shows that 
\begin{equation}
D_{m}=n^{m}\int_{0}^{2\pi }\frac{d\phi }{2\pi }\prod_{i=1}^{m}\left[ 1+x\cos
(\mathbf{k\cdot r}_{i}+\phi )\right]   \label{mFock}
\end{equation}
where it is assumed that $m<<N$.

In order to derive this result, we invert Eq.\ (\ref{phasestate}). Multiply
both sides by $e^{-i\phi (N-n)}=e^{-i\phi N_{b}}$ and integrate over $\phi $
to give 
\begin{equation}
\left| N_{a},N_{b}\right\rangle =\frac{g^{N/2}}{\gamma ^{N_{b}}}\sqrt{\frac{%
N_{a}!N_{b}!}{N!}}\int_{0}^{2\pi }\frac{d\phi }{2\pi }e^{-i\phi N_{b}}\left|
\phi ,N\right\rangle   \label{PhaseSum}
\end{equation}
Thus, by Eq.\ (\ref{eigen}), we find 
\begin{equation}
c_{\mathbf{r}_{m}}\cdots c_{\mathbf{r}_{1}}\left| N_{a},N_{b}\right\rangle
\sim \int_{0}^{2\pi }\frac{d\phi }{2\pi }e^{-i\phi N_{b}}\prod_{i=1}^{m}A(%
\mathbf{r}_{i},\phi )\left| \phi ,N-m\right\rangle   \label{FtoPS}
\end{equation}
We will return to analyze this interesting state below. First consider the $m
$-body Fock correlation function: 
\begin{eqnarray}
D_{m} &=&\left\langle N_{a},N_{b}\right| c_{\mathbf{r}_{m}}^{\dagger }\cdots
c_{\mathbf{r}_{1}}^{\dagger }c_{\mathbf{r}_{m}}\cdots c_{\mathbf{r}%
_{1}}\left| N_{a},N_{b}\right\rangle  \\
&\sim &\int_{0}^{2\pi }\frac{d\phi ^{\prime }}{2\pi }\int_{0}^{2\pi }\frac{%
d\phi }{2\pi }e^{-i\left( \phi -\phi ^{\prime }\right)
N_{b}}\prod_{i=1}^{m}A^{*}(\mathbf{r}_{i},\phi ^{\prime })A(\mathbf{r}%
_{i},\phi )\left\langle \phi ^{\prime },N-m\right| \left| \phi
,N-m\right\rangle 
\end{eqnarray}
Phase states are not actually orthogonal, but for large $N$ they are
essentially so as we show in Appendix A, so, if $m\ll N,$ we can write $%
\left\langle \phi ^{\prime },N-m\right| \left| \phi ,N-m\right\rangle \sim
\delta (\phi -\phi ^{\prime })$ and 
\begin{equation}
D_{m}\sim \int_{0}^{2\pi }\frac{d\phi }{2\pi }\prod_{i=1}^{m}\left| A(%
\mathbf{r}_{i},\phi )\right| ^{2}
\end{equation}
just as we claimed in Eq.\ (\ref{mFock}). We will show how to get this
result by direct calculation, as done in Ref.\ \onlinecite{FLHidden}, in
Appendix B.

This result Eq.\ (\ref{mFock}) has the form of Eq.\ (\ref{mPhase}) but with
the unknown phase integrated over. That makes sense in that our initial Fock
state did not have a phase defined and can, indeed, itself be expressed as a
sum over phase states as in Eq.\ (\ref{PhaseSum}).

Suppose we have started with a Fock state and have made $m-1$ particle
measurements and have found particles at positions $\mathbf{R}_{1},\cdots ,%
\mathbf{R}_{m-1}$ and are about to measure the $m$th. Then the probability
of finding the $m$th particle at position $\mathbf{r}_{m}$ is 
\begin{equation}
P_{m}=n^{m}\int_{0}^{2\pi }\frac{d\phi }{2\pi }g_{m}(\phi )\left[ 1+x\cos (%
\mathbf{k\cdot r}_{m}+\phi )\right]  \label{mthprob}
\end{equation}
where 
\begin{equation}
g_{m}(\phi )=\prod_{i=1}^{m-1}\left[ 1+x\cos (\mathbf{k\cdot R}_{i}+\phi
)\right]  \label{gphi}
\end{equation}
As we show by direct simulation, $g(\phi )$ develops a sharp peak at some 
\textit{a priori} unpredictable phase. If one makes measurements from the
first to the $m$th particle by this prescription, the peak becomes narrower
as one proceeds. Of course, as more measurements are made, the particle
number in each condensate becomes less certain (as, for example, in Eq.\ (%
\ref{Nlesscert})), so phase can be more sharply defined.

Now look back at Eq.\ (\ref{FtoPS}). After a fair number $m$ of
measurements, the real part of $\prod_{i=1}^{m}A(\mathbf{r}_{i},\phi )$
peaks up sharply at some $\phi $ value, call in $\phi _{0}$. That means that
the measurements have converted the Fock wave function into a narrow sum of
phase states around $\left| \phi _{0},N-m\right\rangle .$ The more
measurements that are made, the better is the definition of the phase state.
Measurements in a Fock state provide a way to prepare a phase state. One can
understand the MIT experiments\cite{Ketterle} in this way. The starting
state was prepared as two separate condensates, whose particle numbers could
have been known; many subsequent particle measurements sharpened the phase
to some random value and the final overall observation showed that phase.

\section{NUMERICAL SIMULATION}

We use Eqs. (\ref{mthprob}) and (\ref{gphi}). One choses the initial
position $\mathbf{r}_{1}$ randomly, and then the next particle is chosen
from the probability distribution $P_{2}$ and so on. We will find that, if $m
$ is large enough, $g_{m}$ peaks up at some \textit{a priori} unpredictable
phase angle, $\phi _{0}$ which may fluctuate somewhat as $m$ changes but
gradually settles down. Starting a new experiment from the Fock state will
lead to a randomly different phase angle. We will consider only the case in
which the initial Fock state has $N_{a}=N_{b}=N/2,$ i.e., $x=1.$ It is
convenient to Fourier expand $g_{m}.$ That is, we write 
\begin{equation}
g_{m}(\phi )=a_{0}+\sum_{q=1}^{\infty }\left[ a_{q}\cos q\phi +b_{q}\sin
q\phi \right] 
\end{equation}
In the integration of Eq.\ (\ref{mthprob}) only $a_{0},$ $a_{1,}$ and $b_{1}$
will contribute and doing the integrals gives 
\begin{equation}
P_{m}(\mathbf{r}_{m})\sim 1+\frac{a_{1}}{2a_{0}}\cos (\mathbf{k\cdot r}_{m})-%
\frac{b_{1}}{2a_{0}}\sin (\mathbf{k\cdot r}_{m})  \label{Prob}
\end{equation}

If we define $\cos (\phi _{m})\equiv a_{1}/\sqrt{a_{1}^{2}+b_{1}^{2}}$, $%
\sin (\phi _{m})\equiv b_{1}/\sqrt{a_{1}^{2}+b_{1}^{2}},$ and $A_{m}\equiv 
\sqrt{a_{1}^{2}+b_{1}^{2}}/2a_{0},$ we can write 
\begin{equation}
P_{m}=K[1+A_{m}(\cos \mathbf{k\cdot r}_{m}\cos \phi _{m}-\sin \mathbf{k\cdot %
r}_{m}\sin \phi _{m})]=K\left( 1+A_{m}\cos (\mathbf{k\cdot r}_{m}+\phi
_{m})\right) .  \label{Prob2}
\end{equation}
where $K$ is a normalization factor, and 
\begin{equation}
\tan (\phi _{m})=\frac{b_{1}}{a_{1}}.
\end{equation}
gives the the value of an angle in the $m$th experiment.  Since $P_{m}$ is a
probability, we must have $A_{m}<1$ so that $P_{m}$ is always positive.
Since $A_{m}$ has that property we can write $A_{m}\equiv \sin (\alpha _{m})$
where $0<\alpha _{m}<\pi $ behaves like a polar angle. Then the emerging
phase actually has a space angle designation ($\alpha _{m},\phi _{m}).$ We
will find numerically that $A_{m}\rightarrow 1$ rapidly as we make
measurements. In that case the probability of Eq.\ (\ref{Prob2}) looks just
like the density prediction of Eq.\ (\ref{SimpInt}) and moreover since $%
g_{m}(\phi )$ is a narrow function peaking at $\phi _{0},$ the phase defined
by the Fourier coefficients is the same as that defined by the peak of $%
g_{m},$ as seen using Eq. (\ref{mthprob}).

We work in just one dimension. To choose from the probability of Eq.~(\ref
{Prob}), we need to find the cumulative probability $C_{m}(x)=%
\int_{0}^{x}dx^{\prime }P_{m}(x^{\prime })$ and then solve the equation $%
R=C_{m}(x)$ for $x,$ where $R$ is a random number uniformly distributed on $%
0 $ to $1.$ We take the box size $L=1$ and choose a $k$ value such that $kL$
is an integer times 2$\pi $ to provide periodic boundary conditions. The
normalization of the probability in Eq.(\ref{Prob2}) is just the factor $%
K=1/L=1.$

Figs. 1 and 2 are plots of $\phi _{m}$ and $A_{m}$ versus iteration number
in a particular run of 200 interations; each of these is found from the
amplitudes $a_{0},$ $a_{1},$ $b_{1}$ at each iteration. There is no reason
why $A_{m}$ should be unity from the outset. However, $A_{m}$ does always
proceed to unity after a small number of iterations. The result is that $%
\phi _{m}$ approaches some sharply-defined random phase angle as predicted.
Of course, for small $m$ the fluctuations are relatively large and settle
down only after many measurements. This corresponds to an initially wide
distribution, $g_{m}(\phi )$, which, however, progressively narrows as more
information is gathered. Fig.~3 plots the final angular distribution $%
g_{200}(\phi )$; it is indeed sharply peaked at the same value found from
the iteration limit. Fig.~4 plots the final probability distribution, Eq.~(%
\ref{Prob2}), versus postion $x$ and also shows a histogram of the positions
found in the 200 iterations on the same scale. One sees that these positions
really do fall in the given distribution with the expected oscillations and
with the same phase as gotten in the two other ways. 

\section{DISCUSSION.}

We have shown how one might rigorously treat the interference of two Bose
condensates. The usual assumption of two condensates with individually known
phases gets involved with the thorny questions about whether one can
usefully define the phase of a single condensate. However, using such states
does lead directly to the usual fomulas for interference patterns based on
very simple assumptions. This procedure remains unsatisfying since it may
not be very obvious to the experimenter how to prepare such phase states
before looking at the interference. Experimentally it seems to make no
difference, since without special preparation experiments, even with Fock
states, we have seen how an interference pattern arises using Bose
condensates. The discussion of Sec.\ IV shows explicitly why such
preparation was not necessary. Even if one starts with a state where the
particle number in each cloud is precisely known and many particles are
involved in the measurement, one finds a perfect interference pattern
emerging, with a well-defined relative phase. Starting from a state having a
definite number of particles, the experiment will end up with a state having
a quite definite value of the relative phase. Thus this procedure actually
provides a method of preparing the phase state discussed in Sec.\ III.
Starting from a Fock state, make, say, two hundred position measurements to
get to a narrow $g_{200}(\phi );$ the phase of the wave function of the
remaining state of $N-100$ total particles will be well-defined. The final
result is likely a phase state with a known \emph{total} number of particles
such as that discussed at Eq.\ (\ref{phasestate}), but an unknown number in
each condensate.

In theoretical superfluid calculations it is simpler to treat the problem
with definite phases than to use a Fock state. However, the actual existence
of such a broken symmetry state is subject to question.\cite{LS}$^{-}$\cite
{Johnston} In a ferromagnet one actually has broken the symmetry of the
various directions of magnetization from the presence of some small external
field. The existence of a field that would make $\left\langle \hat{\psi}%
\right\rangle $ non-zero is not so clear, since such states do not conserve
particle number. The existence of a well-defined relative phase established
by measuring particle positions can be established without worrying about
broken symmetry.

If a person feels uncomfortable with the idea of a phase \emph{emerging}
from the series of measurements done on particle position, then he or she
might assume, with no change in theoretical prediction, that that phase
pre-existed within the clouds of particles. That is, the individual
condensates had some relative phase, before they met, as a so-called
``hidden variable'' and the experiments simply bring out this previously
hidden phase. Of course, in the next realization of the experiment, starting
again from a Fock state, the phase will surely emerge with a randomly
different value, in accordance with conventional quantum mechanics, which
expresses the Fock state as a sum over all phase states as given in Eq.\ (%
\ref{PhaseSum}).

\vspace{5mm} \appendix {APPENDIX A. }COHERENT STATES \vspace{5mm}

Consider the following wave function to describe a single momentum state $k,$
which is made up of a mixture of states of known particle number $N_{k}.$ 
\begin{equation}
\left| \alpha _{k}\right\rangle =e^{-\frac{1}{2}\gamma _{k}^{2}}\sum_{N_{k}}%
\frac{\alpha _{k}^{N_{k}}}{\sqrt{N_{k}!}}\left| N_{k}\right\rangle .
\label{Eq1}
\end{equation}
This function is properly normalized. The parameter $\alpha _{k}$ is complex
and we separate it into magnitude $\gamma _{k}$ and phase $\phi _{k}$
according to the notation 
\begin{equation}
\alpha _{k}=\gamma _{k}e^{i\phi _{k}}.
\end{equation}
We can easily compute the average number of particles $\bar{N}_{k}$ in this
state. Let $a_{k}$ be the destruction operator for particles in state $%
\left| N_{k}\right\rangle .$ Then 
\begin{equation}
\bar{N}_{k}=\left\langle \alpha _{k}\right| a_{k}^{\dagger }a_{k}\left|
\alpha _{k}\right\rangle =e^{-\gamma _{k}^{2}}\sum_{N_{k}}\frac{\gamma
_{k}^{2N_{k}}}{N_{k}!}N_{k}=\gamma _{k}^{2}
\end{equation}
and $\gamma _{k}=\left| \alpha _{k}\right| =\sqrt{\bar{N}_{k}}$ .

The state $\left| \alpha _{k}\right\rangle $ has the nice property that it
is an eigenstate of the lowering operator $a_{k}:$%
\begin{equation}
a_{k}\left| \alpha _{k}\right\rangle =e^{-\gamma _{k}^{2}}\sum_{N_{k}}\frac{%
\alpha _{k}^{N_{k}}}{\sqrt{N_{k}!}}a_{k}\left| N_{k}\right\rangle =\alpha
_{k}e^{-\gamma _{k}^{2}}\sum_{N_{k}}\frac{\alpha _{k}^{N_{k}-1}}{\sqrt{%
\left( N_{k}-1\right) !}}\left| N_{k}-1\right\rangle =\alpha _{k}\left|
\alpha _{k}\right\rangle .
\end{equation}
Thus $a_{k}$ has a non-zero expectation value in this state: 
\begin{equation}
\left\langle \alpha _{k}\right| a_{k}\left| \alpha _{k}\right\rangle =\alpha
_{k}=\sqrt{\bar{N}_{k}}e^{i\phi _{k}}
\end{equation}
Clearly the states $\left| \alpha _{k}\right\rangle $ provide a definite
phase. Of course they are then not eigenstates of the number operator.

Next construct a multi-level many-body state with many $k$ values possible.
This takes the form $\left| \alpha _{k_{1}},\alpha _{k_{2},}\alpha
_{k_{3}}\cdots \right\rangle =\left| \alpha _{k_{1}}\right\rangle \left|
\alpha _{k_{2}}\right\rangle \left| \alpha _{k_{3}}\right\rangle \cdots $%
\begin{equation}
\left| \left\{ \alpha _{k}\right\} \right\rangle \equiv \left| \alpha
_{k_{1}},\alpha _{k_{2},}\alpha _{k_{3}}\cdots \right\rangle =\sum_{\left\{
N_{k}\right\} }\prod_{k}\left[ e^{-\frac{1}{2}\gamma _{k}^{2}}\frac{\alpha
_{k}^{N_{k}}}{\sqrt{N_{k}!}}\right] \left|
N_{k_{1}},N_{k_{2}},N_{k_{3}}\cdots \right\rangle 
\end{equation}
where $\left\{ N_{k}\right\} $ means sum over all possible numbers of
particles $N_{k_{i}}$ in all the $k$-states.

With such a state, we can consider the expectation value of the full field
operator $\hat{\psi}(\mathbf{r}).$ Expand the field operator in plane wave
states as

\begin{equation}
\hat{\psi}(\mathbf{r})=\sqrt{\frac{1}{V}}\sum_{k}e^{i\mathbf{k}\cdot \mathbf{%
r}}a_{k}  \label{field exp}
\end{equation}
where $V$ is the volume of the system. We have 
\begin{equation}
\left\langle \left\{ \alpha _{k}\right\} \right| \hat{\psi}(\mathbf{r}%
)\left| \left\{ \alpha _{k}\right\} \right\rangle =\sum_{k}\sqrt{\frac{\bar{N%
}_{k}}{V}}e^{i\phi _{k}}e^{i\mathbf{k}\cdot \mathbf{r}}.
\end{equation}
If one of the $k$-states is macroscopically occupied, say, the momentum
state $k=k_{a},$ then we can write 
\begin{equation}
\left\langle \hat{\psi}(\mathbf{r})\right\rangle =\sqrt{\bar{n}_{a}}e^{i%
\mathbf{k}_{a}\cdot \mathbf{r}}e^{i\phi _{a}}+\xi
\end{equation}
where $n_{a}=N_{a}/V$ and $\xi $ is the total contribution of the
non-condensed states. The leading term $\psi _{a}(\mathbf{r})=\sqrt{\bar{n}%
_{a}}e^{i\mathbf{k}_{a}\cdot \mathbf{r}}e^{i\phi _{a}}$ represents a
condensate wave function having a definite phase $\phi _{a}$ but non-sharp
number of particles.

Consider the case of the interference of a double condensate in momentum
states $k_{a}$ and $k_{b}.$ With coherent states having only these two
momentum states occupied we can write 
\begin{equation}
\left| \alpha _{a}\alpha _{b}\right\rangle =e^{-\frac{1}{2}\left( \gamma
_{a}^{2}+\gamma _{b}^{2}\right) }\sum_{N_{a}N_{b}}\frac{\alpha
_{a}^{N_{a}}\alpha _{b}^{N_{b}}}{\sqrt{N_{a}!N_{b}!}}\left|
N_{a},N_{b}\right\rangle   \label{2CondPhaseSt}
\end{equation}
where the averages $\bar{N}_{a}=\gamma _{a}^{2}$ and $\bar{N}_{b}=\gamma
_{b}^{2}$ are both macroscopic quantities. We manipulate the sums slightly
in terms of particle creation operators $a^{\dagger }$ and $b^{\dagger }$
for the two states. If $N=N_{a}+N_{b}$ then we have 
\begin{eqnarray}
\left| \alpha _{a}\alpha _{b}\right\rangle  &=&e^{-\frac{1}{2}\left( \gamma
_{a}^{2}+\gamma _{b}^{2}\right) }\sum_{N,N_{a}}\frac{\alpha
_{a}^{N_{a}}\alpha _{b}^{N-N_{a}}}{\sqrt{N_{a}!\left( N-N_{a}\right) !}}%
\left| N_{a},N-N_{a}\right\rangle   \nonumber \\
&=&e^{-\frac{1}{2}\left( \gamma _{a}^{2}+\gamma _{b}^{2}\right)
}\sum_{N,N_{a}}\frac{1}{N_{a}!\left( N-N_{a}\right) !}\left( \alpha
_{a}a^{\dagger }\right) ^{N_{a}}\left( \alpha _{b}b^{\dagger }\right)
^{N-N_{a}}\left| 0\right\rangle   \nonumber \\
&=&e^{-\frac{1}{2}\left( \gamma _{a}^{2}+\gamma _{b}^{2}\right) }\sum_{N}%
\frac{1}{N!}\left( \alpha _{a}a^{\dagger }+\alpha _{b}b^{\dagger }\right)
^{N}\left| 0\right\rangle   \nonumber \\
&=&e^{-\frac{1}{2}\left( \gamma _{a}^{2}+\gamma _{b}^{2}\right) }e^{\left(
\alpha _{a}a^{\dagger }+\alpha _{b}b^{\dagger }\right) }\left|
0\right\rangle 
\end{eqnarray}

The second last form of Eq.\ (\ref{Expform}) is the ``phase state'' used in
Sec. IV. 
\begin{equation}
\left| \alpha _{a}\alpha _{b};N\right\rangle \sim \left( \alpha
_{a}a^{\dagger }+\alpha _{b}b^{\dagger }\right) ^{N}\left| 0\right\rangle .
\label{phiN}
\end{equation}
We see that it is a substate of the more general coherent state.

\vspace{5mm} \appendix {APPENDIX B. NEAR-ORTHOGONALITY OF PHASE STATES} 
\vspace{5mm}

We calculate the inner product of two phase states to show that they are
nearly orthogonal for large particle number. From Eq. (\ref{phasestate}) we
find 
\begin{equation}
\left\langle \phi ^{\prime },N\right. \left| \phi ,N\right\rangle =\frac{1}{%
g^{N}}\sum_{n=1}^{N}\frac{N!}{n!(N-n)!}\left[ \gamma ^{2}e^{i\left( \phi
-\phi ^{\prime }\right) }\right] ^{N-n}=\frac{1}{g^{N}}\left[ 1+\gamma
^{2}e^{i\left( \phi -\phi ^{\prime }\right) }\right] ^{N}
\end{equation}
This is a very sharply peaked function of $\left( \phi -\phi ^{\prime
}\right) .$ To see this, Taylor expand the logarithm of this in powers of $%
\left( \phi -\phi ^{\prime }\right) $ and then exponentiate the result
keeping only terms to $\left( \phi -\phi ^{\prime }\right) ^{2}$ . The
result is 
\begin{equation}
\left\langle \phi ^{\prime },N\right. \left| \phi ,N\right\rangle =\exp
\left[ -\frac{N}{2(1+\gamma ^{2})^{2}}\left( \phi ^{\prime }-\phi \right)
^{2}\right] \exp \left[ -i\frac{N}{(1+\gamma ^{2})}\left( \phi ^{\prime
}-\phi \right) \right] 
\end{equation}
In the limit of very large $N,$ this is proportional to a delta function of
the $\left( \phi -\phi ^{\prime }\right) $ as we assumed in the discussion
of Sec. IV.

\vspace{5mm} \appendix {APPENDIX C. ALTERNATIVE DERIVATION OF THE $D_{m}$
EQUATION} \vspace{5mm}

We want to derive the general expression of Eq.\ (\ref{mFock}) for the
correlation function $D_{m}.$ Consider this quantity in its original form
for a Fock state: 
\begin{eqnarray}
D_{m} &=&\left\langle N_{a},N_{b}\right| c_{\mathbf{r}_{m}}^{\dagger }\cdots
c_{\mathbf{r}_{1}}^{\dagger }\cdots c_{\mathbf{r}_{m}}\cdots c_{\mathbf{r}%
_{1}}\left| N_{a},N_{b}\right\rangle   \nonumber \\
&=&\frac{1}{V^{m}}\left\langle N_{a},N_{b}\right| \left( a^{\dagger }+e^{-i%
\mathbf{k}\cdot \mathbf{r}_{m}}b^{\dagger }\right) \cdots \left( a^{\dagger
}+e^{-i\mathbf{k}\cdot \mathbf{r}_{1}}b^{\dagger }\right) \cdots   \nonumber
\\
&&\times \left( a+e^{i\mathbf{k}\cdot \mathbf{r}_{m}}b\right) \cdots \left(
a+e^{i\mathbf{k}\cdot \mathbf{r}_{1}}b\right) \left|
N_{a},N_{b}\right\rangle 
\end{eqnarray}
Because this is a diagonal matrix element in Fock space, each time an $a$
occurs, there must be a matching $a^{\dagger }.$ Similarly for the $b$
operators. We are assuming $m\ll N_{a}$ or $N_{b}$ so that we can always
write $a\left| N_{a}-p,N_{b}-l\right\rangle \approx $ $\sqrt{N_{a}}\left|
N_{a}-p-1,N_{b}-l\right\rangle ,$ etc. Thus each $a^{\dagger }a$ gives $%
N_{a},$ and each $b^{\dagger }b$ an $N_{b}$. Consider a particular
combination product: 
\begin{eqnarray}
\left( a^{\dagger }+e^{-i\mathbf{k}\cdot \mathbf{r}_{j}}b^{\dagger }\right)
\left( a+e^{i\mathbf{k}\cdot \mathbf{r}_{l}}b\right)  &=&a^{\dagger
}a+b^{\dagger }b+a^{\dagger }be^{i\mathbf{k}\cdot \mathbf{r}_{l}}+b^{\dagger
}ae^{-i\mathbf{k}\cdot \mathbf{r}_{j}}  \nonumber \\
&\rightarrow &N_{a}+N_{b}+\sqrt{N_{a}N_{b}}e^{i\mathbf{k}\cdot \mathbf{r}%
_{l}}+\sqrt{N_{a}N_{b}}e^{-i\mathbf{k}\cdot \mathbf{r}_{j}}
\end{eqnarray}
with the restriction that every time an $e^{i\mathbf{k}\cdot \mathbf{r}_{l}}$%
-type term occurs there \emph{must} be a corresponding $e^{-i\mathbf{k}\cdot 
\mathbf{r}_{j}}$-type term somewhere in the overall product to give the
proper balance of creation and destruction operators. Thus one gets a series
of terms of the form $F_{q_{m}}(\mathbf{r}_{m})F_{q_{m-1}}(\mathbf{r}%
_{m-1})\cdots F_{q_{2}}(\mathbf{r}_{2})F_{q_{1}}(\mathbf{r}_{1})$ where 
\begin{eqnarray}
F_{0}(\mathbf{r}_{i}) &=&N_{a}+N_{b}  \nonumber \\
F_{\pm 1}(\mathbf{r}_{i}) &=&\sqrt{N_{a}N_{b}}e^{\pm i\mathbf{k}\cdot 
\mathbf{r}_{l}}
\end{eqnarray}
and the sum of all the $q_{j}$ vanishes. That is, we have 
\begin{equation}
D_{m}=\frac{1}{V^{m}}\sum_{\{q\}}F_{q_{m}}\cdots F_{q_{2}}F_{q_{1}}
\end{equation}
where $\{q\}$ means sum on all $q_{i}$ with the restriction that $%
\sum_{i}q_{i}=0.$

The restriction on the $q$ values can be lifted if we insert the integral 
\begin{equation}
\int_{0}^{2\pi }\frac{d\phi }{2\pi }e^{i\phi \sum q_{i}}=\delta _{\sum
q_{i},0}
\end{equation}
which allows us to write 
\begin{eqnarray}
D_{m} &=&\frac{1}{V^{m}}\int_{0}^{2\pi }\frac{d\phi }{2\pi }
\prod_{i=1}^{m}\left[ F_{0}(\mathbf{r}_{i})+e^{i\phi }F_{1}(\mathbf{r}%
_{i})+e^{-i\phi }F_{-1}(\mathbf{r}_{i})\right]  \nonumber \\
&=&n^{m}\int_{0}^{2\pi }\frac{d\phi }{2\pi }\prod_{i=1}^{m}\left[ 1+x\cos
\left( \mathbf{k}\cdot \mathbf{r}_{i}+\phi \right) \right]
\end{eqnarray}
as we wished to prove.

\newpage

\begin{figure}[h]
\centering
\includegraphics[width=5in, height=3.69in]{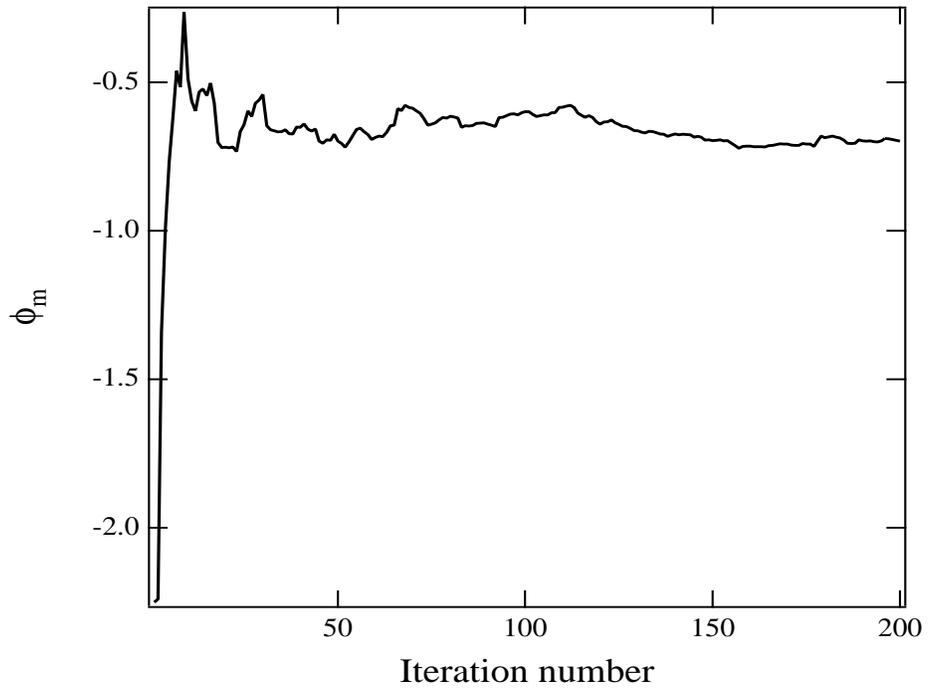}
\caption{The phase angle $\phi _{m}$ as a function of interation step.}
\label{Phi}
\end{figure}
\begin{figure}[h]
\centering
\includegraphics[width=5in, height=3.20in]{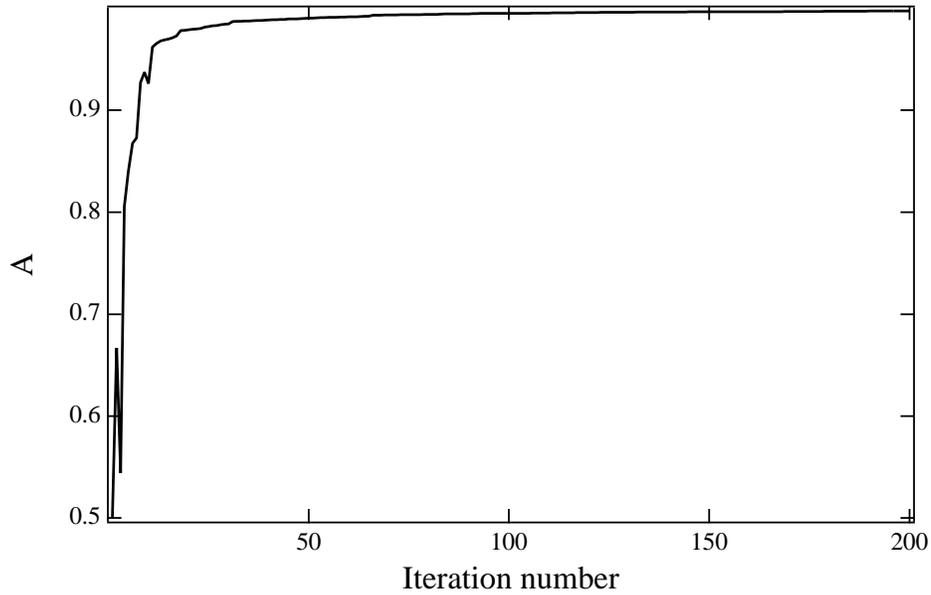}
\caption{The amplitude $A=\sin \alpha $ as a function of interation step.
This always proceeds to unity.}
\label{Amp}
\end{figure}
\begin{figure}[h]
\centering
\includegraphics[width=5in, height=2.81in]{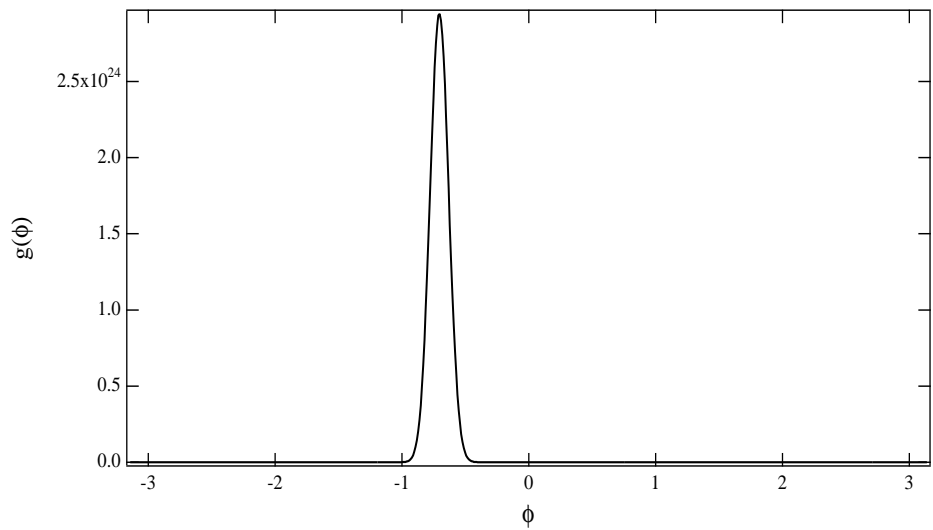}
\caption{The angular distribution $g(\phi )$ as a function of angle. This
peaks at the same phase angle as given in Fig.~1.}
\label{gg}
\end{figure}
\begin{figure}[h]
\centering
\includegraphics[width=5in, height=5.89in]{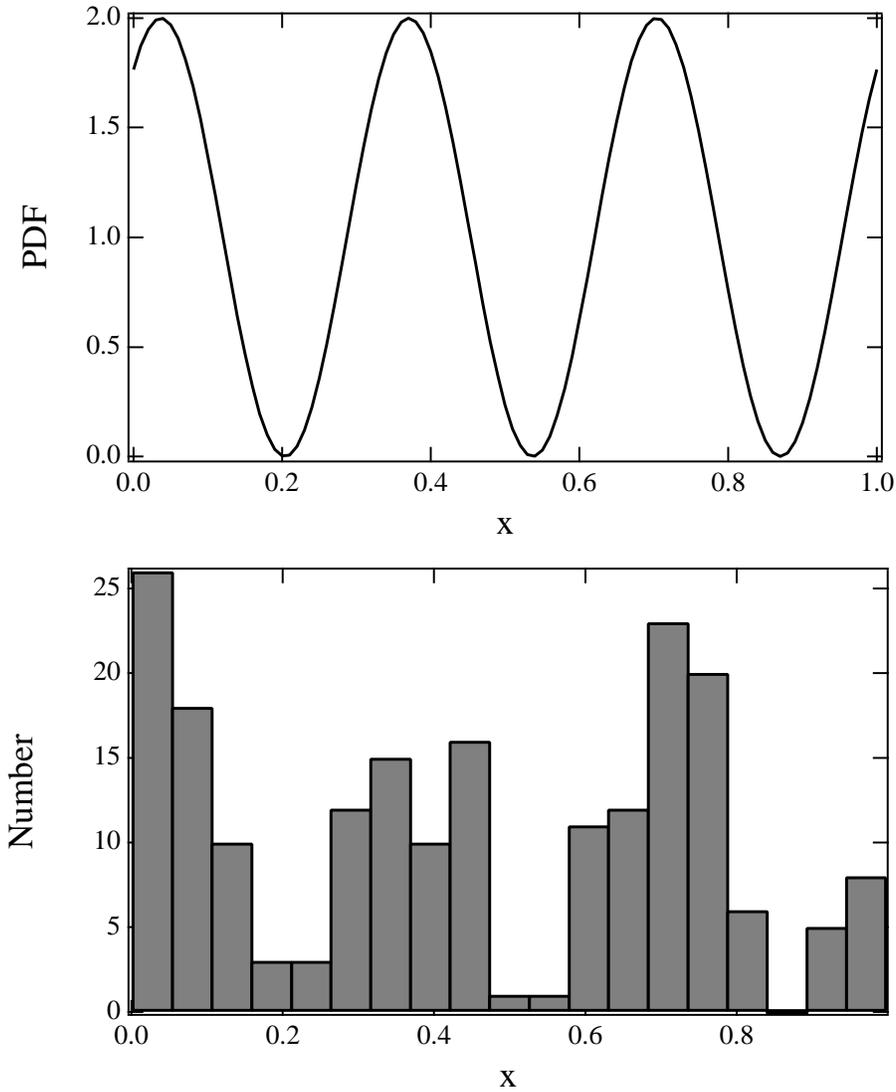}
\caption{The probability distribution function for position in the
interference pattern as calculated and the histogram of this as found in
simulated experiments. The phase is found here to be the same as in the
other approaches.}
\label{pdf}
\end{figure}

\end{document}